\documentclass{article}
\usepackage[left=0.75in, right=1in, top=1in, bottom=1in]{geometry}
\usepackage{cite,graphicx,subfigure,setspace}
\usepackage[small,bf,up]{caption}

\begin{document}
\title{Photoluminescence from silicon dioxide photonic crystal cavities with embedded silicon nanocrystals}
\author{Yiyang Gong$^{*,1}$, Satoshi Ishikawa$^{2}$, Szu-Lin Cheng$^{3}$, Yoshio Nishi$^{1}$, and Jelena Vu\v{c}kovi\'{c}$^{1}$ \\
	\small\textit{$^{1}$ Department of Electrical Engineering, Stanford University, Stanford, CA 94305} \\
	\small\textit{$^{2}$ Corporate Manufacturing Engineering Center, Toshiba Corporation, Yokohama, 235-0017, Japan} \\
	\small\textit{$^{3}$ Department of Material Science and Engineering, Stanford University, Stanford, CA 94305} \\
	\small\textit{*email:yiyangg@stanford.edu}}
\begin{@twocolumnfalse}
\maketitle
\begin{abstract}
One dimensional nanobeam photonic crystal cavities are fabricated in silicon dioxide with silicon nanocrystals. Quality factors of over $9\times 10^3$ are found in experiment, matching theoretical predictions, with mode volumes of $1.5(\lambda/n)^3$. Photoluminescence from the cavity modes is observed in the visible wavelength range 600-820 nm. Studies of the lossy characteristics of the cavities are conducted at varying temperatures and pump powers. Free carrier absorption effects are found to be significant at pump powers as low as a few hundred nanowatts.
\end{abstract}
\end{@twocolumnfalse}
The promise of integrating electronics and optics has fueled the search for efficient light emitting materials compatible with Silicon complementary metal-oxide-semiconductor (CMOS) processing. The system of silicon nanocrystals (Si-NCs) embedded in a silicon dioxide (SiO$_{2}$, or silica) host has been proposed as one such candidate, capable of emitting in the visible and near-infrared wavelengths and possibly exhibiting gain \cite{SiNC_book,Pavesi_SiNC}. While the quantum confinement properties of the material relax some constraints for radiative recombination in the indirect band gap silicon, the efficiency of this material is still low. The photonic crystal (PC) cavity system, with high quality ($Q$-) factor and low mode volume ($V_{m}$), could enhance radiative emission from the nanocrystals via Purcell enhancement of the radiative rate, which is proportional to $Q/V_{m}$. By enhancing the radiative rate with respect to the non-radiative decay rates, the efficiency of the Si-NC emission can be increased. In addition, the high-$Q$ cavity could be used to spectrally filter signals and define operating wavelengths from the broad luminescence range of the Si-NCs. The lossy mechanisms of this material coupled to micro-disks ($\mu$-disks) were previously analyzed \cite{Rohan_FCA,Brongersma_udiskloss}, and it was noted that free carrier absorption may play a significant role in lossy processes and may ultimately inhibit gain in this material \cite{Rohan_FCA}. In this work, we demonstrate Si-NCs embedded in a silicon dioxide host coupled to high $Q$ and low $V_{m}$ PC nanobeam cavities, fabricated by CMOS compatible processing. In addition, we investigate the temperature and pump power dependence of the cavities, and compare them to previous work.  We observe that free carrier absorption remains significant in high-$Q$ and low $V_{m}$ photonic crystal cavities, and is even present at very low pump powers of a few hundred nanowatts.

\begin{figure}[hbtp]
\centering
\includegraphics[width=3.2in]{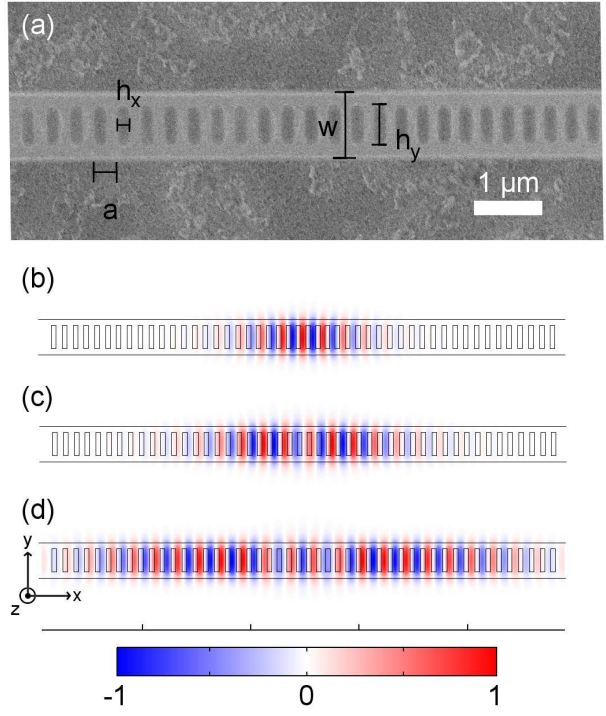}
\caption{(a) SEM picture of the fabricated beam structure. The $E_{y}$ component of the (b) first- (c) second- and (d) third-order TE-like modes supported by the nanobeam with parameters $w=3.2a$, $h_{x}=0.5a$, $h_{y}=0.7w$, and thickness $d=0.7a$ are also shown.}
\label{fig:modes}
\end{figure}

Si-NCs can be grown in a SiO$_{2}$ host by either Si implantation \cite{Pavesi_SiNC} or plasma enhanced chemical vapor deposition (PECVD). One of the previous hindrances to producing high $Q$ PC cavities was the small optical band gap afforded by low index materials in a two dimensional (2D) PC setting. For example, Barth et al. were able to experimentally demonstrate $Q=3,400$ in silicon nitride (Si$_{3}$N$_{4}$) \cite{Benson_SiNhet} with a heterostructure two dimensional PC cavity in a triangular lattice. While a full photonic band gap in a 2D photonic crystal is difficult to achieve in low index materials, one dimensional (1D) nano-beam cavities that rely on a one-dimensional band gap in the direction parallel to the beam and total internal reflecion in the remaining two directions perpendicular to the beam can lead to high $Q$-factors with low index contrast. Recent work has found that such a design can be applied to materials with a variety of indices of refraction, including Si ($n=3.5$) \cite{MIT_1D,Marko_Si1D}, Si$_{3}$N$_{4}$ ($n=2.0$) \cite{Painter_1Dmodes,Marko_SiN1D}, and even SiO$_{2}$ ($n=1.46$) \cite{Gong_quartz1D}. In particular, the experimentally demonstrated $Q$s exceed $10^5$ for Si \cite{Marko_Si1D}, $10^4$ for Si$_{3}$N$_{4}$ \cite{Painter_1Dmodes}, and $10^3$ for SiO$_{2}$ \cite{Gong_quartz1D}, while maintaining $V_{m} < 2.0 (\lambda/n)^3$.

\begin{figure}[hbtp]
\centering
\includegraphics[width=3.2in]{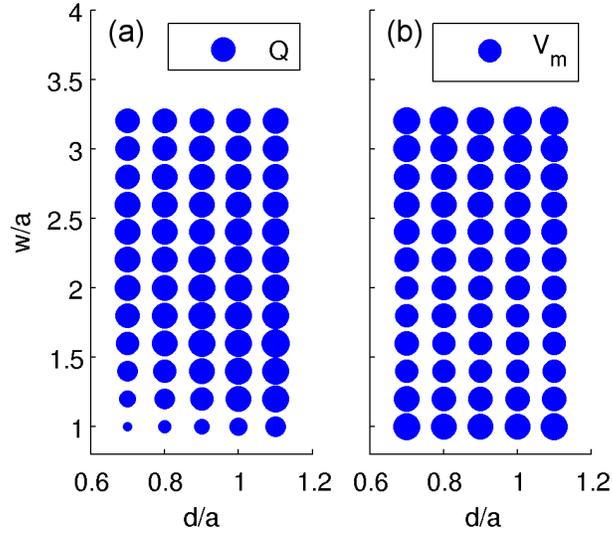}
\caption{(a) $Q$ and (b) $V_{m}$ for the TE$_{0}$ mode for different beam widths ($w$) and thicknesses ($d$). We keep the same air hole design for the simulated cavities. The reference points in the legend represent $Q=20,000$ and $V_{m}=1.1(\lambda/n)^3$.}
\label{fig:QV}
\end{figure}

We employ the parabolic potential well design of previous works \cite{Painter_1Dmodes,Gong_quartz1D}, where the distance between air holes on the beam increases parabolically from the center of the beam outward. The cavity extends 6 air holes on either side of the center of the cavity, where the distance between air holes is 0.9$a$ at the center of the cavity, while holes outside of the cavity in the photonic crystal mirror have with lattice constant $a$. The beam has width ($w$) and thickness ($d$), while the air holes on the beam have horizontal size ($h_{x}=0.5a$) and vertical size ($h_{y}=0.7w$) (Fig. \ref{fig:modes}(a)). We simulate the cavity using the three dimensional finite difference time domain (3D-FDTD) method, with a discretization of 20 units per lattice constant and apply perfectly matched layer absorbing boundary conditions. For a cavity with $d/a=0.7$ and $w/a=3.2$, and assuming the Si-NC doped oxide material to have an index of refraction of $n=1.7$, we find that the cavity supports at least three transverse electric (TE-) like modes, which have dominant $E_{y}$ field profiles shown in Fig \ref{fig:modes}(b)-(d), and are referred to as the first (TE$_{0}$, or fundamental), second (TE$_{1}$), and third order (TE$_{2}$) modes, respectively \cite{Painter_zipper}. The TE$_{0}$, TE$_{1}$, and TE$_{2}$ modes have normalized mode frequencies of $a/\lambda=$ $0.131$, $0.125$, and $0.121$, respectively, with the electric field energy of higher order modes occupying regions on the beam with increasing lattice constants. In addition, we compute the $Q$ of the cavity as $Q=\omega U/P$, where $\omega$ is the frequency of the mode, $U$ is the total energy of the mode, and $P$ is the time averaged radiated power. We find that the TE$_{0}$, TE$_{1}$, and TE$_{2}$ modes have $Q$s of $2.0\times 10^4$, $1.3\times 10^4$, and $2.9\times 10^3$, and mode volumes of 1.6$(\lambda/n)^3$, 2.5$(\lambda/n)^3$, and 4.4$(\lambda/n)^3$, respectively, for a reference index of $n=1.7$. 

\begin{figure}[hbtp]
\centering
\includegraphics[width=3.2in]{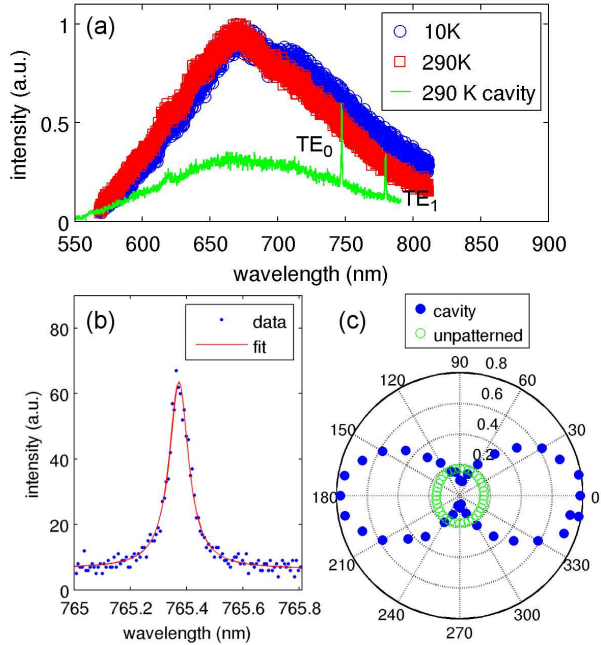}
\caption{(a) PL spectra from unpatterned film at 10 K and 290 K, as well as a cavity spectrum at 290 K with the first two TE modes visible. (b) The PL spectrum of a representative fundamental cavity mode at 290 K, and the fit to a Lorentzian with $Q=9,000$. (c) The polarization angle dependence of the cavity mode at 290 K, along with the angle dependence of PL from an unpatterned region. 0$^{\circ}$ corresponds to the $y$-direction of Fig. \ref{fig:modes}(d).}
\label{fig:cavity}
\end{figure}

We also vary the width ($w$) and thickness ($d$) of the beam with the same fixed air hole design, and find $Q$ and $V_{m}$ for different beam parameters (Fig \ref{fig:QV}(a) and (b), respectively. For $w<2.4a$, we find that the confinement of the mode (and thus the $Q$) increases as the beam increases in either width or thickness, and that the $V_{m}$ correspondingly increases. Such a trade-off between $Q$ and $V_{m}$ was previously observed in silica nanobeam cavities \cite{Gong_quartz1D}, and is also present in other types of cavities, including 2D PC cavities and $\mu$-disk systems. In addition, we observe that for all beam widths, such a trade-off is maintained as the thickness of the beam is increased. However, we see that the $Q$ of the cavity mode saturates as the width of the beam is increased beyond $w>2.4a$, as confinement of the mode in the $y$-direction is no longer dominant in the overall confinement of the mode. Indeed, the thicknesses ($d$) of the beams are far smaller than the width at $w=2.4a$, and increasing $d$ further increases confinement. Likewise, the $k$-space emission profile could remain largely unchanged by the increase of the beam width, thus limiting the $Q$ achieved by this air hole design. In the future, modifications to the air hole design could increase the $Q$s for this type of cavity. In order to maximize the Purcell enhancement for this cavity, we would maximize the $Q/V_{m}$ ratio, which (in this parameter space), is achieved at $w=1.6a$ and $d=1.1a$, corresponding to $Q=25,000$ and $V_{m}=1.1(\lambda/n)^3$. This represents an order of magnitude increase in $Q$ and 7 times reduction in mode volume compared to $\mu$-disks with a similar thickness and a diameter of 5 $\mu$m \cite{Brongersma_udiskloss}.

The cavities are fabricated by a fully CMOS compatible process with similar parameters as in simulations. In particular, the Si-NC doped oxide layer is grown on top of a silicon substrate by PECVD at 350$^{\circ}$C in a N$_{2}$ atmosphere with a gas flow of SiH$_{4}$:N$_{2}$O=1400:300sccm. The growth rate is approximately 46 nm/min, and the final oxide thickness is 200 nm. The sample is then annealed in a N$_{2}$ atmosphere at 900$^{\circ}$C for 1 hr, followed by a forming gas (95\% N$_{2}$, 5\% H$_{2}$) anneal at 500$^{\circ}$C for 1 hr. The refractive index of the layer is measured as $n=1.7$. Next, e-beam lithography is performed with a 250 nm layer of ZEP-520A as the resist. After development of the resist, the pattern is transferred to the oxide layer with a CHF$_{3}$:O$_{2}$ (100 sccm/2 sccm ratio) chemistry dry etch \cite{Gong_quartz1D}. Finally, the beam is undercut with a 1.0 Torr XeF$_{2}$ dry etch, removing approximately 3$\mu$m of silicon under the oxide layer. The final fabricated structure is shown in Fig. \ref{fig:modes}(a). We vary the lattice constant $a$ in fabrication to create cavities with a variety of wavelengths, where all beams have a width of $w=3.2a$ (due to fabrication constraints).

\begin{figure}[hbtp]
\centering
\includegraphics[width=3.2in]{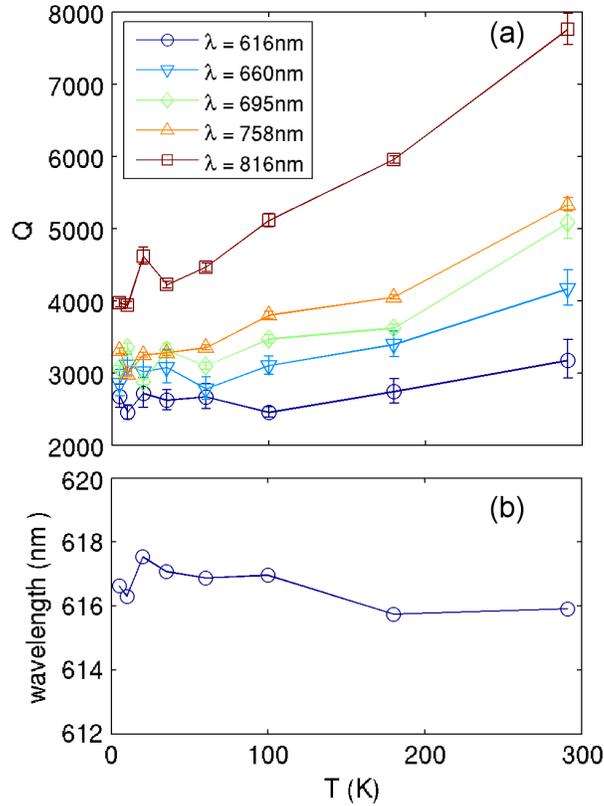}
\caption{(a) The $Q$s of cavities at different wavelengths as a function of temperature. (b) The wavelength of an example cavity as a function of temperature.}
\label{fig:tempcavs}
\end{figure}

The micro-photoluminescence ($\mu$-PL) experiments are conducted with a 75$\times$ objective lens with numerical aperture $NA = 0.75$. The sample is pumped from normal incidence ($z$) with either a continuous-wave (CW) 400 nm laser diode or a frequency doubled Ti:Sapph laser at 390 nm with 3 ps pulses at 13 ns intervals. The photoluminescence is also collected from normal incidence through the same objective and directed to a Si CCD array in a spectrometer. The cryogenic temperature experiments are conducted in a helium flow cryostat, with temperatures as low as 5 K. Room temperature data was taken outside of vacuum. We observe the PL from an unpatterned region of the sample at both 10 K and 290 K, using the 400 nm CW laser as the pump (Fig. \ref{fig:cavity}(a)). The PL spectrum is centered at around 670 nm, and is the same at both temperatures, which suggests that the change in the homogeneous linewidth for the Si-NCs is small with temperature. Finally, we observe that the PL of the sample pumped with the pulsed Ti:Sapph laser (not shown) was the same as the PL in Fig \ref{fig:cavity}(a) at both temperatures, similar to previous work \cite{Vergnat_SiNClife}.

We also investigate the cavities in the same $\mu$-PL configuration, first at room temperature. A representative cavity spectrum at 290 K is shown in Fig. \ref{fig:cavity}(a), where at least the first two ordered TE modes (TE$_{0}$ and TE$_{1}$) are visible. We also plot the spectrum of the fundamental mode of a high-$Q$ cavity along with a fit to a Lorentzian lineshape, representing $Q=9,000$, in Fig \ref{fig:cavity}(b). Finally, we measure the polarization angle dependence of the cavity mode by placing a polarizer and a half waveplate in the PL collection path, plotted in Fig. \ref{fig:cavity}(c). The cavity mode is dominated by the $E_{y}$ field, and the mode is measured to have a linear polarization in the $y$-direction. On the other hand, the PL collected from the unpatterned region is unpolarized. As seen from the FDTD simulations, the TE$_{0}$ mode has the highest $Q$ and lowest $V_{m}$. Thus, we choose to work with the fundamental mode for the remainder of this work to maximize the Purcell effect.

We measure the $Q$-factors of various cavities throughout the Si-NC PL spectrum at temperatures between room temperatures and 5 K, as shown in Fig \ref{fig:tempcavs}(a). The cavities are all pumped with the CW diode laser at very low pump powers ($200$ nW), which is necessary to reduce the cavity losses stemming from free carrier absorption (FCA). The different cavities have different lattice constant $a$, but have the same fixed $w/a=3.2$ ratio, $d=200$ nm, and the same air hole design.  We find that for all cavities, the cavity $Q$ continuously drops to approximately one half of the room temperature value at 5 K. Such change in cavity $Q$ can be attributed to the difference in the homogeneous linewidth of the Si-NCs at the two temperatures. As seen in previous work on coupling Er emission to PC cavities, the Purcell enhancement of absorption is degraded when the homogeneous linewidth of the emitter far exceeds that of the cavity linewidth \cite{Erbcav_gain}. Single Si-NCs have measured linewidths of over 100 nm at room temperature, but have far narrower linewidths down to 1 nm at 35 K \cite{Linnros_SiNCwidth}. Thus, the narrower linewidth at low temperature increases the absorption of the Si-NCs coupled to the cavity mode, and lowers the $Q$ of the cavities. The temperature dependence of $Q$ suggests that the homogeneous linewidth of the Si-NCs continuously decreases with temperature, much as in previous work \cite{Linnros_SiNCwidth}. We also measure the cavity wavelength as a function of temperature, shown for an example cavity in Fig. \ref{fig:tempcavs}(b). There is a only a small shift of the cavity wavelength from 5 K to room temperature, which is due to systematic movements of the sample in the cryostat with varying temperatures.

\begin{figure}[hbtp]
\centering
\includegraphics[width=3.2in]{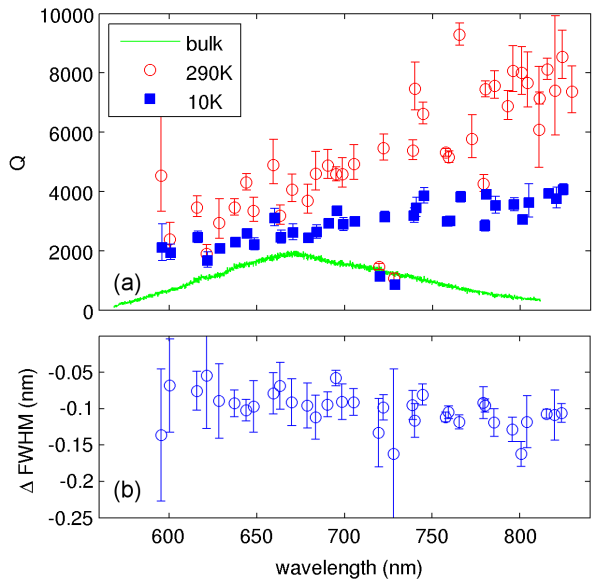}
\caption{(a) The $Q$s of the one set of cavities vs. the cavity wavelengths, at 10 K and 290 K. The PL from an unpatterned region of the sample is shown for reference. (b) The change in the cavity linewidth as the temperature is increased from 10 K to 290 K.}
\label{fig:cavities}
\end{figure}

In addition, we study the $Q$s of many cavities at the extremes of the temperature at 10 K and room temperature, where the same set of cavities was measured at both temperatures (Fig. \ref{fig:tempcavs}(a)). We again notice that the $Q$s of the cavities are higher at room temperature than the $Q$s of the same cavities at 10 K. Because the lattice constant increases for the longer wavelength cavities, the $d/a$ ratio decreases with increasing cavity wavelength. We note in Fig. \ref{fig:tempcavs}(a) that the cavities have increasing $Q$ at longer wavelengths, which disagrees with the simulated trend of higher $Q$s for thicker beams (higher $d/a$). While this discrepancy could be caused by higher sensitivity to surface roughness for cavities operating at shorter wavelengths, we do not observe this effect in similar cavities made in pure SiO$_{2}$ \cite{Gong_quartz1D}. Thus, we attribute the decreasing $Q$ with decreasing lattice constant to the material absorption of the Si-NCs, which increases with decreasing wavelength (as more NCs contribute to absorption). We also measure the change in the cavity linewidth for the same set of cavities between 290 K and 10 K, shown in Fig. \ref{fig:cavities}(b). We notice that the decrease in the linewidth with increasing temperature is approximately uniform and equal to 0.10-0.15 nm for the entire wavelength range of the Si-NC PL. The losses in the cavity can be related to the cavity $Q$ by:
\begin{equation}
\frac{\omega}{Q} = \frac{\omega}{Q_{0}} + \alpha(T,P)
\end{equation}
where $\omega$ is the cavity frequency, $Q_{0}$ is the intrinsic cavity $Q$, and $\alpha(T,P)$ is the pump power and temperature dependent loss rate. The intrinsic $Q$ chnage with temperature is negligible, as a result of the small change in refractive index (as can be conluded form the negligible change in cavity wavelength shift with temperature, shown in Fig. \ref{fig:tempcavs}(b)). In this case, the measured linewidth difference is the change in $\alpha(T,P)$ with temperature, and suggests that the linewidth narrowing of the Si-NCs is uniform throughout the PL spectrum.

\begin{figure}[hbtp]
\centering
\includegraphics[width=3.5in]{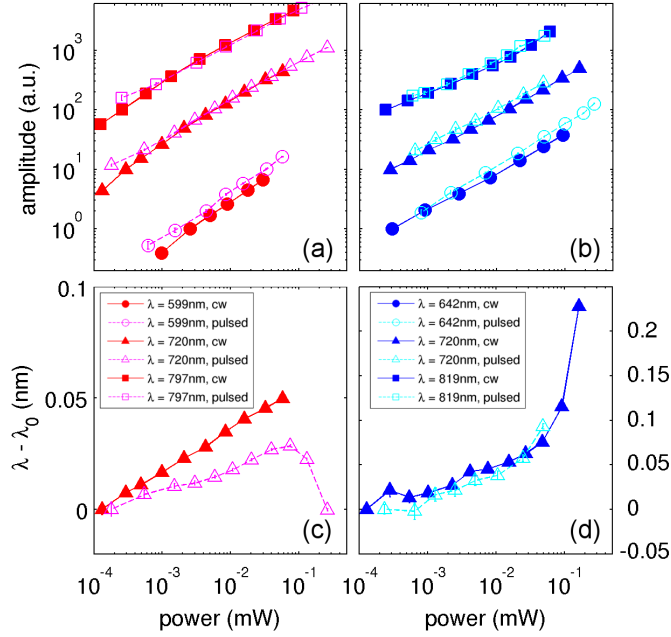}
\caption{The pump power dependence of the integrated intensity for different cavities at (a) 290 K and (b) 10 K. The amplitude traces of each cavity are offset by factors of 10 to allow clear viewing. The pump power dependence of the change in cavity wavelength (with respect to the low pump power wavelength, $\lambda_{0}$) for different cavities at (c) 290 K and (d) 10 K. Power dependences were taken with both the CW diode laser and the pulsed frequency doubled Ti:Sapph. The pump power is measured in front of the ojective.}
\label{fig:powercavs1}
\end{figure}

We also use 10 K and 290 K as representative temperatures for measuring the power dependence of the cavities. We employ both the CW diode laser and the pulsed doubled Ti:Sapph laser in the power dependence studies, and we pump various representative cavities throughout the PL wavelength range. By fitting the obtained spectra to Lorentzian lineshapes, we obtain the integrated emission, cavity wavelength, and cavity linewidth. In Fig \ref{fig:powercavs1}(a)-(b), we first investigate the integrated cavity intensity obtained from the fits at 290 K and 10 K, respectively, as a function of pump power. We observe that the cavity output is approximately linear in the pump power, with some slight sub-linear characteristics, mostly likely due to FCA. In addition, we find that at both temperatures, the output for the pulsed pump generates slightly more PL. Finally, at room temperature, the output intensities of the cavities have slightly lower slope for the pulsed pump than for the CW pump. On the other hand, the slopes of the light-in light-out curves are approximately the same for the different types of pump at 10 K. Such sub-linear behavior at small pump powers is similar to the trend from previous work with Si-NCs coupled to $\mu$-disk cavity modes \cite{Rohan_FCA}. The smaller increases in the cavity amplitude with increasing pump power for pulsed pumping suggest that lossy mechanisms are more readily found with pulses and high instantaneous pump intensities.

We also find the pump dependence of the change in the cavity wavelength from the fits to a Lorentzian lineshape, and plot the results for the cavities around 720 nm in Fig. \ref{fig:powercavs1}(c)-(d), obtained at 290 K and 10 K, respectively. We observe that in both cases, the cavity continuously red-shifts with increasing pump power, implying the thermo-optic effect as the cause. At room temperature, the pulsed pump generates far less heating than the CW pump, while at 10 K, the wavelength shifts are approximately the same between the two pump methods. In addition, at high pump powers for both CW and pulsed pumping, we observe melting of the cavities, where the wavelength of the cavities abruptly blueshifts (e.g. in Fig. \ref{fig:powercavs1}(c) for the pulsed pumping). In order to avoid this regime, pump powers are kept below 500 $\mu$W for all experiments. We also note that the wavelength shift for the cavities can be related to the index of refraction change caused by the thermo-optic effect by:
\begin{equation}
\Delta\lambda \approx -\lambda \frac{\Gamma}{2}\frac{\Delta\epsilon}{\epsilon},
\end{equation}
where $\Gamma$ is the fraction of the mode overlapping with the Si-NC doped oxide, $\lambda$ is the unperturbed cavity wavelength, $\Delta \lambda$ is the change in cavity wavelength, $\Delta\epsilon$ is the change in dielectric constant, and $\epsilon$ is the dielectric constant of the Si-NC doped oxide. For the TE$_{0}$ cavity mode at 720 nm, we obtain $\Gamma=0.52$ from FDTD simulations, and find that the index of refraction shift is $\Delta n=2.2\times 10^{-4}$, which corresponds to a temperature change of 20 K, using the bulk thermal-optic coefficient of oxide, $dn/dT=1\times 10^{-5}/$K. Such a change in index (and temperature) is achieved with approximately two orders of magnitude less power than in the $\mu$-disk setting \cite{Rohan_FCA}. This large change is due to the lack of heat conduction from the nanobeam cavities, as noted from previous work with thermo-optical bistability in 1D nanobeam cavities fabricated in silicon \cite{Notomi_optbi}. The changes in index with temperature are doubled when pumping the cavities at 10 K, possibly due to reduced air convection inside a vacuum.

\begin{figure}[hbtp]
\centering
\includegraphics[width=3.2in]{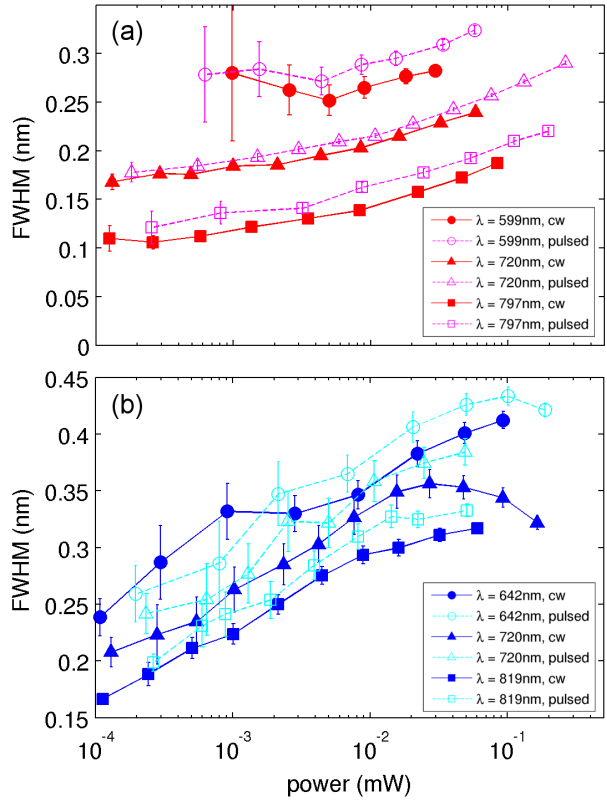}
\caption{The linewidths of representative cavities as the pump power is changed at (a) 290 K and (b) 10 K. Both a CW diode laser and a doubled mode-locked Ti:Sapph laser are used as the pump source. The pump power is measured in front of the ojective.}
\label{fig:powercavs}
\end{figure}

Finally, we measure the cavity linewidth for representative cavities throughout the Si-NC PL spectrum, at both 290 K and 10 K in Fig. \ref{fig:powercavs}(a) and (b), respectively. We note that the cavity linewidth continuously increases with increasing pump power, regardless of cavity wavelength or cavity $Q$. The linewidth data resembles the linewidth data for $\mu$-disk modes in Ref. \cite{Rohan_FCA}, though the low pump power $Q$s for the nanobeam cavities are approximately 4-8 times higher. Such broadening of the cavity linewidth is thought to be resulting from FCA. Taking the pump beam spot area to be 4 $\mu$m$^2$, we find that the power necessary to observe FCA is more than an order of magnitude less than in $\mu$-disks, corresponding to a pump flux of 3 W/cm$^2$ \cite{Rohan_FCA}. Following the slope of the curves in Fig. \ref{fig:powercavs}, smaller powers could also generate losses from free carrier absorption. As described previously \cite{Rohan_FCA}, the onset of FCA only occurs when the FCA dominates other processes such as Mie scattering and band-to-band absorption in nanocrystals. On the other hand, in high $Q$, low $V_{m}$ systems, we observe that free-carrier absorption dominates the other processes, and that linewidth broadening effect from FCA occurs at even small powers. In addition, the linewidth increases until at high pump powers ($> 100 \mu$W), there is a sudden increase due to overheating and melting of the cavities. In previous work, the losses arising from FCA in $\mu$-disks were saturated when the carrier density is sufficiently high to promote Auger recombination \cite{Rohan_FCA}. However, we do not reach that regime, as the pump intensities required are higher than that to destroy the cavities.  We further note that the cavity linewidth for each cavity is narrower with the CW pump than with the pulsed pump at room temperature. Such evidence corroborates the conclusion that pulsed pumping induces more losses in the cavities than CW pumping.

\begin{figure}[hbtp]
\centering
\includegraphics[width=3.6in]{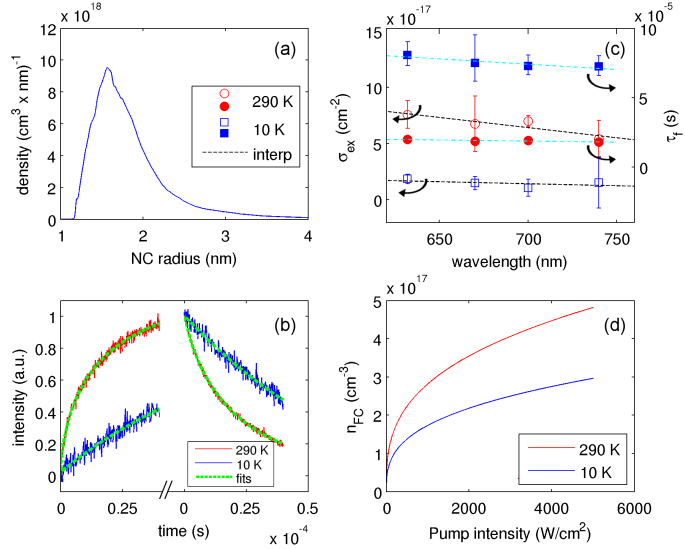}
\caption{(a) The Si-NC density distrubution of this sample as a function of Si-NC radius. (b) Time resolved measurements of the Si-NC rise and fall time. Fits to extended exponential distributions yield a rise time of 17 $\mu$s and a fall time of 20 $\mu$s for the sample at 290 K, and a rise time of 78 $\mu$s and a fall time of 79 $\mu$s for the sample at 10 K. (c) The excitation cross section and the lifetimes ($\tau_{f}$) of the Si-NCs as a function of emission wavelength, for both 290 K and 10 K. (d) The calculated free carrier concentration as a function of pump intensity for 290 K and 10 K.}
\label{fig:FCA}
\end{figure}

We attempt to characterize the FCA mechanism following the procedure in Ref \cite{Rohan_FCA}. First, we find the distribution of Si-NC sizes by assuming that the energy of emission for the Si-NCs is related to the size of the NCs by \cite{Rohan_FCA}:
\begin{equation}
\hbar \omega = E_{g,Si} + \frac{3.73}{(2R)^{1.39}},
\end{equation}
where $E_{g,Si}=1.12$eV is the bandgap energy of bulk Si, $\hbar \omega$ is the energy of emission in eV, and $R$ is the radius of the NC in nanometers. The emission at each energy is proportional to the density of nanocrystals at a particular size, and the proportionality constant is found by finding the volume ratio of Si in the entire film, assuming complete phase segregation. By finding the mass fractions of Si ($x_{Si}=0.42$) and O ($x_{O}=0.47$) obtained from x-ray photoelectron spectroscopy (XPS), we find that the distribution of nanocrystal sizes ($\rho(R)$) as the distribution shown in Fig. \ref{fig:FCA}(a). As expected, the distribution of NCs has smaller radii than than the NCs of Ref. \cite{Rohan_FCA}, as the NC emission has been pushed to shorter wavelengths. By integrating the distribution for all radii, we find a total nanocrystal density of $N_{NC}=7\times 10^{18}$ cm$^{-3}$, which is comparable to the densities obtained in other works \cite{Pavesi_SiNC,Rohan_FCA}.

We also find the pump power depedent rise ($\tau_{r}$) and fall ($\tau_{f}$) times of the Si-NCs, by chopping the CW diode pump with an acousto-optic modulator with 15 ns rise and fall times. Representative time traces from unpatterned regions of the sample can be found in Fig. \ref{fig:FCA}(b), for the sample at 290 K and 10 K. Extended exponential functions are fitted to the data to find $\tau_{r}$ and $\tau_{f}$. The $\tau_{f}$ of the sample is found to be approximately 20 $\mu$s at 290 K, and approximately 80 $\mu$s at 10 K (Fig. \ref{fig:FCA}(c)). The longer lifetime at low temperatures is expected, as similar samples of porous silicon also demonstrate a dramatic increase in lifetime with decreasing temperature, which was attributed to longer radiative lifetimes and higher fraction of the emission coming from the diffusion of carriers \cite{Pavesi_PorousSi}. The excitation cross section of the Si-NCs at a particular wavelength $\lambda$ ($\sigma_{ex}(\lambda)$) can be related to $\tau_{r}$, $\tau_{f}$, and the pump flux $\phi$ by \cite{Rohan_FCA}:
\begin{equation}
\frac{1}{\tau_{r}(\lambda)} = \sigma_{ex}(\lambda) \phi + \frac{1}{\tau_{f}(\lambda)}.
\end{equation}
By conducting time-resolved measurements at various wavelengths, using 10 nm band-pass filters to spectrally filter the emission, we find $\sigma_{ex}$ by a linear fit of the pump flux dependent difference of $1/\tau_{r}-1/\tau_{f}$ to $\phi$, with the results shown in Fig. \ref{fig:FCA}(c). Finally, the occupation (the number of electron-hole pairs) of individual Si-NCs at a particular size and pump flux ($f(R,\phi)$) is found by the relation \cite{Rohan_FCA}:
\begin{equation}
\frac{\sigma_{ex}(\lambda) \phi}{\hbar \omega_{p}} - \frac{f(R,\phi)}{\tau_{f}(\lambda)} - \frac{f^3(R,\phi)}{\tau_{A}(\lambda)} = 0,
\end{equation}
where $\omega_{p}$ is the pump laser frequency, and $\tau_{A}$ is the Auger recombination time. The Auger time constant can be found by $\tau_{A} = 1/C_{A}\times (V/2)^3$, where $C_{A} = 4\times 10^{-31}$ cm$^{6}$/s is the Auger recombination coefficient for bulk Si, and $V$ is the volume of the Si-NC assuming a spherical particle shape \cite{Maly_SiNCtime,Rohan_FCA}. Although the bulk Si Auger coefficient is used, the Auger recombination time of Si-NCs with similar sizes as the ones in this work has been observed to match well with Auger time constant found when using the bulk figure \cite{Maly_SiNCtime}. Finally, the density of free carriers at a pump flux $\phi$ ($n_{FC}(\phi)$) can be found by integrating the occupation for NCs of all sizes \cite{Rohan_FCA}:
\begin{equation}
n_{FC}(\phi)=\int_R \! f(R,\phi) \rho(R) \, dR.
\end{equation}
We plot the results in Fig. \ref{fig:FCA}(d), and observe that due to the smaller excitation cross-section at 10 K, the free carrier concentration at 10 K at any particular pump flux is approximately one half that of the free carrier concentration at 290 K.

\begin{figure}[hbtp]
\centering
\includegraphics[width=3.6in]{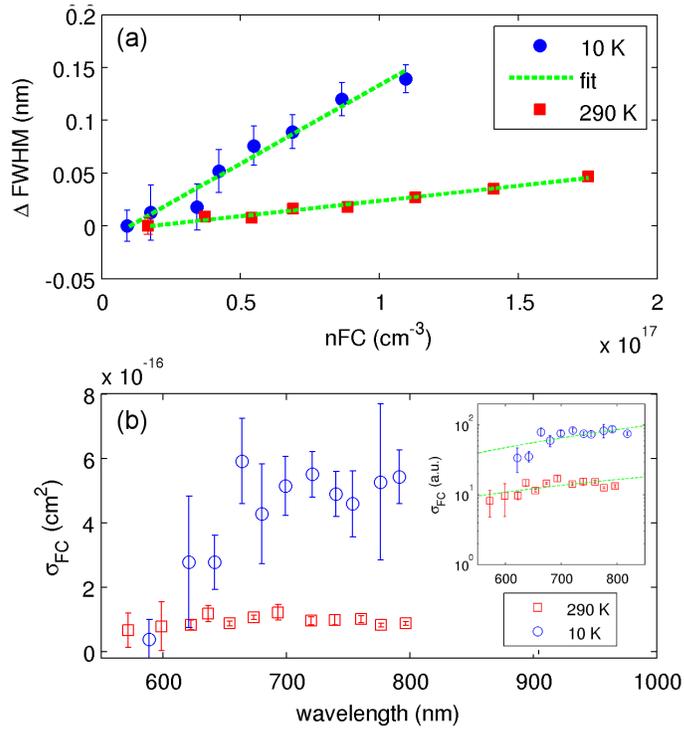}
\caption{(a) The change in cavity linewidth as a function free carrier density for representative cavities at 290 K and 5 K. Linear fits to the data are also shown.  (b) The free carrier absorption cross-section ($\sigma_{FCA}$) obtained from linear fits such as those in part (a), as a function of wavelength, for different cavities throughout the PL spectrum of the Si-NCs. The inset shows the same data renormalized in a small $V_{m}$ setting. The dashed lines represent fits to a $\lambda^b$ model, where $b=1.3$ and $1.8$ for the 290 K and 6 K data, respectively.}
\label{fig:cavFCA}
\end{figure}

Finally, this free carrier concentration can be related to the change in linewidth of the cavity modes, assuming the FCA is the dominant absorption process:
\begin{equation}
\Delta \mbox{FWHM} = \frac{\lambda^2 \Gamma}{2 \pi n_{eff}}\sigma_{FCA}(\lambda) n_{FC},
\end{equation}
where $\Delta$FWHM is the change in the cavity linewidth, $\Gamma$ is the overlap of the cavity mode with the active material, $n_{eff}$ is the effective index of the cavity mode, and $\sigma_{FCA}(\lambda)$ is the free carrier absorption cross-section. We calculate $n_{eff}$ and $\Gamma$ from the FDTD simulations with various beam dimensions. We plot the change in linewidth (obtained from the power dependence curves) against $n_{FC}$, and find $\sigma_{FCA}$ through a linear fit to the slope (Fig. \ref{fig:cavFCA}(a)) for cavities at both 290 K and 10 K. In addition, we plot all of the fitted $\sigma_{FCA}$ for cavities throughout the Si-NC PL spectrum at both temperatures, as a function of the cavity wavelength, in Fig. \ref{fig:cavFCA}(b). We first note that the mean free carrier absorption obtained in the nanobeam cavities is a factor of 4 increased from that obtained in Ref. \cite{Rohan_FCA} at room temperature. In addition, as noted earlier, we obtain similar changes in the cavity linewidth with increasing pump power at 290 K and 10 K. However, because the free carrier density at the same pump power is lower at 10 K than at 290 K, we calculate that the $\sigma_{FCA}$ is approximately 6 times higher at 10 K than at 290 K (Fig. \ref{fig:cavFCA}(b)). However, we do not observe the $\lambda^2$ dependence of $\sigma_{FCA}$ generally associated with FCA\cite{Rohan_FCA}, especially at room temperature, where $\sigma_{FCA}$ is seen to decrease with wavelength.

While the calculation above allows easy direct comparison with the figures obtained in Ref. \cite{Rohan_FCA}, they assume that the optical mode of the cavity is traveling in an effective medium with some effective index and group velocity. While such an assumption is valid for large cavities such as $\mu$-disks, they are oversimplified for the case of high $Q$, low $V_{m}$ cavities. In the cavity setting, the Purcell enhanced strength of absorption is given by:
\begin{equation}
\alpha \propto \int_\omega |g_{0}|^2 \rho(\omega) d\omega \propto \int_\omega \rho(\omega)/V_{m} d\omega,
\end{equation}
where $|g_{0}|^2\propto 1/V_{m}$ is the emitter-field interaction term, and $\rho(\omega)$ is the joint density of states associated with the electronic transition for FCA and the optical density of states of the cavity. However, since the transition for FCA is a broad continuum, the normalized narrow bandwidth optical density of states from the cavity is integrated out, resulting in no enhancement of the absorption rate in high $Q$ cavities. Thus, in this case, the Purcell enhanced absorption should only be proportional to $\Gamma$ and the inverse of $V_{m}$. We renormalize the calculated $\sigma_{FC}$  of Fig. \ref{fig:cavFCA}(b) with those considerations and plot the results in the inset of the same figure, along with their respective fits to a $\sigma_{FC} \propto \lambda^{b}$ model. We obtained $b=1.3$ and $1.8$ for the 290 K and 10 K cases, respectively, which is more consistent with FCA being the cause of the power dependent absorption.

In conclusion, we have demonstrated high $Q$ nanobeam cavities in low-index Si-NC doped SiO$_{2}$, with experimental $Q$ as high as 9,000, and mode volumes of $1.5(\lambda/n)^3$, operating in the visible wavelengths from 600 nm to 820 nm. We have also observed a decrease in $Q$-factor with decreasing temperature, which is associated with enhanced absorption losses from the Si-NCs with decreasing temperature \cite{Erbcav_gain}. Nevertheless, high $Q$s were maintained for all temperatures and all wavelengths. Finally, we have also investigated the free carrier absorption mechanism associated with the pump power dependent linewidth broadening, and computed the free carrier absorption cross-section. As shown in previous work \cite{Rohan_FCA}, the free carrier absorption cross-section in Si-NCs are much larger than the corresponding figure in bulk Si. In this work, we have shown that the absorption cross-section of the Si-NCs is further increased when the NCs are placed in a high $Q$, low $V_{m}$ cavity, as we demonstrate a 4-fold increase in the measured cross-section at room temperature. In addition, we also measure the free carrier absorption at cryogenic temperatures, and find that the FCA cross-section at 10 K is increased by nearly one order of magnitude from the value at room temperature. This evidence suggests that cavity enhanced absorption may be applicable to processes such as FCA, and are not limited to just emission of single quantum dots \cite{Dirk_PCQDcontrol}, ensembles of atoms with large homogeneous broadening \cite{Erbcav_gain}, and even Mie scattering \cite{Kippenberg_SiNCtoroid}. Even more, the large effective absorption coefficients of Si-NCs in microcavities could hamper development of a lasing source, whether the Si-NCs serve as the emitters or as the sensitizer as in Si-NCs mediated emission from Er doped SiO$_{x}$. Finally, it is likely that cooling down systems involving Si-NCs would present additional challenges. Although many non-radiative phonon mediated recombination processes decrease with decreasing temperature, we have demonstrated that FCA increases dramatically with decreasing temperature as well, possibly offsetting some of the gains. Nevertheless, these Si-NC beams are made by fully CMOS compatible fabrication techniques, and certainly present interesting platform for applications such as sensors, as their cavity modes have high overlap with the environment and their emission spans a wide range of wavelengths in the visible. Such cavities would also be bio-compatible, and could be functionalized for probing emission from molecules. Finally, the enhancement of emission could be increased by modifications in the cavity design, and creating high $Q$ cavities in low index materials remains an interesting challenge.

The authors would like to acknowledge the MARCO Interconnect Focus Center, the Toshiba corporation, and the NSF graduate research fellowship (YG) for funding. Fabrication was done at Stanford Nanofabrication Facilities.


\begin{thebibliography}{widest-label}
\bibitem{SiNC_book}
	\textit{Towards the First Silicon Laser}, ed. Lorenzo Pavesi, Sergey Gaponenko, and Luca Dal Negro. Kluwer Academic Publishers, Netherlands (2003).
\bibitem{Pavesi_SiNC}
	L. Pavesi, L. Dal Negro, C. Mazzoleni, G. Franz\`o, and F. Priolo. ``Optical gain in silicon nanocrystals," Nature \textbf{408}, 440 (2000).
\bibitem{Brongersma_udiskloss}
	R. D. Kekatpure and M. L. Brongersma, ``Fundamental photophysics and optical loss processes in Si-nanocrystal-doped microdisk resonators," Phys. Rev. A \textbf{78}, 023829 (2008).
\bibitem{Rohan_FCA}
	R. D. Kekatpure and M. L. Brongersma, ``Quantification of Free-Carrier Absorption in Silicon Nanocrystals with an Optical Microcavity," Nano Lett. \textbf{8}, 3787 (2008).
\bibitem{Benson_SiNhet}
	M. Barth, N. N\"{u}sse, J. Stingl, B. L\"{o}chel, and O. Benson, ``Emission properties of high-Q silicon nitride photonic crystal heterostructure cavities," Appl. Phys. Lett. \textbf{93}, 021112 (2008).
\bibitem{MIT_1D}
	J. S. Foresi, P. R. Villeneuve, J. Ferrera, E. R. Thoen, G. Steinmeyer, S. Fan, J. D. Joannopoulos, L. C. Kimerling, H. I. Smith, and E. P. Ippen, ``Photonic-bandgap microcavities in optical waveguides, " Nature \textbf{390}, 143 (1997).
\bibitem{Marko_Si1D}
	P. B. Deotare, M. W. McCutcheon, I. W. Frank, M. Khan, and M. Lon\v{c}ar, ``High Quality factor photonic crystal nanobeam cavities," Appl. Phys. Lett., \textbf{94}, 121106 (2009).
\bibitem{Painter_1Dmodes}
	M. Eichenfield, R. Camacho, J. Chan, K. J. Vahala, and  O. Painter, ``A picogram- and nanometre-scale photonic-crystal optomechanical cavity," Nature, \textbf{459}, 550 (2009).
\bibitem{Marko_SiN1D}
	M. W. McCutcheon and M. Lon\v{c}ar, ``Design of an ultrahigh Quality factor silicon nitride photonic crystal nanocavity for coupling to diamond nanocrystals," Optics Express, \textbf{16}, 19136 (2008).
\bibitem{Gong_quartz1D}
	Y. Gong and J. Vu\v{c}kovi\'{c}, ``Photonic crystal cavities in silicon dioxide," Appl. Phys. Lett., \textbf{96}, 031107 (2010).
\bibitem{Painter_zipper}
	J. Chan, M. Eichenfield, R. Camacho, and O. Painter, ``Optical and mechanical design of a `zipper' photonic crystal optomechanical cavity," Opt. Expr. \textbf{17}, 3802 (2009).
\bibitem{Vergnat_SiNClife}
	H. Rinnert, O. Jambois, and M. Vergnat, ``Photoluminescence properties of size-controlled silicon nanocrystals at low temperatures," J. Appl. Phys. \textbf{106}, 023501 (2009͒).
\bibitem{Erbcav_gain}
	Y. Gong, M. Makarova, S. Yerci, R. Li, M. J. Stevens, B. Baek, S. W. Nam, R. H. Hadfield, S. N. Dorenbos, V. Zwiller, J. Vu\v{c}kovi\'{c}, and L. Dal Negro. ``Linewidth narrowing and Purcell enhancement in photonic crystal cavities on an Er-doped silicon nitride platform," Opt. Expr. \textbf{18}, 2601 (2010).
\bibitem{Linnros_SiNCwidth}
	I. Sychugov, R. Juhasz, J. Valenta, and J. Linnros, ``Narrow Luminescence Linewidth of a Silicon Quantum Dot," Phys. Rev. Lett. \textbf{94}, 087405 (2005).
\bibitem{Notomi_optbi}
	L.-D. Haret, T. Tanabe, E. Kuramochi, and M. Notomi, ``Extremely low power optical bistability in silicon demonstrated using 1D photonic crystal nanocavity," Opt. Expr. \textbf{17}, 21108 (2009).
\bibitem{Pavesi_PorousSi}
	L. Pavesi and M. Ceschini, ``Stretched-exponential decay of the luminescence in porous silicon," Phys. Rev. B \textbf{48}, 17625 (1993).
\bibitem{Maly_SiNCtime}
	F. Troj\'{a}nek, K. Neudert, M. Bittner, and P. Mal\'{y}, ``Picosecond photoluminescence and transient absorption in silicon nanocrystals," Phys. Rev. B, \textbf{72}, 075365 (2005͒).
\bibitem{Dirk_PCQDcontrol}
	D. Englund, D. Fattal, E. Waks, G. Solomon, B. Zhang, T. Nakaoka, Y. Arakawa, Y. Yamamoto, and J. Vu\v{c}kovi\'{c},  ``Controlling the Spontaneous Emission Rate of Single Quantum Dots in a 2D Photonic Crystal," Phys. Rev. Lett. \textbf{95}, 013904 (2005).
\bibitem{Kippenberg_SiNCtoroid}
	T. J. Kippenberg, A. L. Tchebotareva, J. Kalkman, A. Polman, and K. J. Vahala, ``Purcell-Factor-Enhanced Scattering from Si Nanocrystals in an Optical Microcavity,"  Phys. Rev. Lett. \textbf{103}, 027406 (2009).



\end{thebibliography}
\end{document}